\documentclass[fleqn,usenatbib]{mnras}

\usepackage{newtxtext,newtxmath}

\usepackage[T1]{fontenc}

\DeclareRobustCommand{\VAN}[3]{#2}
\let\VANthebibliography\thebibliography
\def\thebibliography{\DeclareRobustCommand{\VAN}[3]{##3}\VANthebibliography}

\usepackage{graphicx}	
\usepackage{amsmath}	
\usepackage{pdflscape}
\usepackage{enumitem}
\usepackage{multicol}

\title[On UDG1137+16]{A Photometric and Kinematic Analysis of UDG1137+16 (dw1137+16): Probing Ultra-Diffuse Galaxy Formation in a Group Environment}

\author[J. S. Gannon et al.]{
Jonah S. Gannon,$^{1}$\thanks{E-mail: jgannon@swin.edu.au}
Bililign T. Dullo,$^{2}$
Duncan A. Forbes,$^{1}$
R. Michael Rich,$^{3}$
Javier Rom\'an,$^{4}$
\newauthor
Warrick J. Couch,$^{1}$
Jean P. Brodie,$^{1,5}$
Anna Ferr\'e-Mateu,$^{6,1}$
Adebusola Alabi,$^{5}$
\newauthor
and Jeremy Mould $^{1}$
\\
$^{1}$Centre for Astrophysics and Supercomputing, Swinburne University, John Street, Hawthorn VIC 3122, Australia\\
$^{2}$ Departamento de F\'isica de la Tierra y Astrof\'isica, Instituto de F\'isica de Part\'iculas y del Cosmos IPARCOS, Universidad Complutense de Madrid, E-28040 Madrid, Spain\\
$^{3}$ Department of Physics \& Astronomy, University of California Los Angeles, 430 Portola Plaza, Los Angeles, CA 90095-1547, USA\\
$^{4}$ Instituto de Astrof\'isica de Andaluc\'ia (CSIC), Glorieta de la Astronom\'ia, 18008 Granada, Spain\\
$^{5}$University of California Observatories, 1156 High Street, Santa Cruz, CA 95064, USA\\
$^{6}$Institut de Ci\'encies del Cosmos (ICCUB), Universitat de Barcelona (IEEC-UB), Barcelona 08028, Spain\\
}

\date{Accepted XXX. Received YYY; in original form ZZZ}

\pubyear{2020}

\begin{document}
\label{firstpage}
\pagerange{\pageref{firstpage}--\pageref{lastpage}}
\maketitle

\begin{abstract}
The dominant physical formation mechanism(s) for ultra-diffuse galaxies (UDGs) is still poorly understood. Here, we combine new, deep imaging from the Jeanne Rich Telescope with deep integral field spectroscopy from the Keck II telescope to investigate the formation of UDG1137+16 (dw1137+16). Our new analyses confirm both its environmental association with the low density UGC 6594 group, along with its large size of 3.3 kpc and status as a UDG. The new imaging reveals two distinct stellar components for UDG1137+16, indicating that a central stellar body is surrounded by an outer stellar envelope undergoing tidal interaction. Both the components have approximately similar stellar masses. From our integral field spectroscopy we measure a stellar velocity dispersion within the half-light radius (15 $\pm$ 4 $\mathrm{km\ s^{-1}}$) and find that UDG1137+16 is similar to some other UDGs in that it is likely dark matter dominated. Incorporating literature measurements, we also examine the current state of UDG observational kinematics. Placing these data on the central stellar velocity dispersion -- stellar mass relation, we suggest there is little evidence for UDG1137+16 being created through a strong tidal interaction. Finally, we investigate the constraining power current dynamical mass estimates (from stellar and globular cluster velocity dispersions) have on the total halo mass of UDGs. As most are measured within the half-light radius, they are unable to accurately constrain UDG total halo masses.
\end{abstract}




\section{Introduction}

With large half-light radii ($R_{e}$ $>$ 1.5 kpc) and low surface brightnesses ($\mu_{0, g}$ $>$ 24 mag $\mathrm{arcsec^{-2}}$) observations of `ultra-diffuse galaxies' (UDGs; \citealp{VanDokkum2015}) have raised multiple scientific questions that are yet to be satisfactorily explained. Prime among these, is the dark matter content, and by extension the total dark matter halo mass, of UDGs. While strongly debated, evidence exists for UDGs exhibiting both a paucity \citep{vanDokkum2018, Danieli2019, ManceraPina2019} and an overabundance \citep{Beasley2016, vanDokkum2017, vanDokkum2019b, Martin-Navarro2019, Forbes2020, Gannon2020} of dark matter. It is clear that despite both fitting the same categorical definition, such objects are likely subject to different formation pathways. 

While some authors suggest UDGs may primarily form external to galaxy clusters \citep{Amorisco2016, Roman2017b, DiCintio2017, Chan2018, Alabi2018, Roman2019} others suggest that the tidal stripping of stellar material in infalling cored dark matter halos can create UDGs \citep{Carleton2018}. In contrast, some suggest tidal effects are subtler with tidal heating increasing the UDGs stellar extent while keeping the stellar material of the galaxy largely intact \citep{Jiang2019, Carleton2018, Sales2019}. In such tidal formation scenarios the mass and density of the dark matter halo play a key role governing the strength of its role in UDG formation \citep{Yozin2015, Carleton2018, Martin2019, Sales2019}. Further investigation of both the total dark matter halo mass in UDGs and its profile (i.e., core vs cusp) is required to better illuminate their formation pathways.

In the tidally formed UDG simulations of \citet{Carleton2018} 13\% of simulated UDGs experienced a pericentric passage recent enough to be expected to display tidal features. Those UDGs known to be associated with tidal features (e.g., VLSB-A \citealp{Mihos2015}, CenA-MM-DW3 \citealp{Crnojevic2016} and NGC1052-DF4 \citealp{Montes2020}) are in contrast to discoveries of large numbers of UDGs that do not exhibit such features (e.g., the catalogues of \citealp{Yagi2016} or \citealp{Alabi2020}). As of yet, there is little evidence for a large population of UDGs displaying tidal features in the most complete sample studied so far located in the Coma Cluster \citep{Mowla2017}. It is difficult to reconcile how external tidal effects can play a role in the creation of UDGs without creating a large population of UDGs with tidal features. One possible explanation is that UDGs reside in massive dark matter halos whose larger mass shields the stellar material from tidal stripping \citep{Mowla2017}. We note however, that the work of \citet{Munoz2008} suggests deeper imaging studies may be required to detect such features leaving this an open line of inquiry.  


\citet{Sales2019} suggested that UDGs born via tidal stripping in a cluster environment should have significantly lower velocity dispersions at fixed stellar mass than those resulting from other formation pathways. Similar predictions, but with smaller effect, are also made by \citet{Carleton2018}. The predicted substantial diversity of UDG velocity dispersions can also explain the substantial variation in UDG mass to light (M/L) ratios \citep{Carleton2018, Sales2019}.

If UDGs are to form external to clusters without their tidal field influencing formation, it is necessary to study their properties in less dense group and field environments. The processing of UDGs in a galaxy group is thought to be critical in some simulations (e.g., \citealp{Martin2019}). Here UDGs are expected to form most of their stellar mass rapidly, early in the Universe coring their dark matter halo \citep{Martin2019}. The cored halo profile makes UDGs more susceptible to tidal heating which both suppresses further star formation and expands the stellar content of the galaxy \citep{Carleton2018, Martin2019}. Other studies have shown similar results for UDGs undergoing an expansion cycle due to stellar feedback that are then accreted into a dense environment, quenching star formation. This combination can `freeze' their large size and slowly decrease their surface brightness due to the passive evolution of the stellar population \citep{DiCintio2017, Chan2018, Tremmel2019}. Indeed, many UDG formation scenarios rely on either the pre-existence or creation of a dark matter core in the halo density profile (e.g., \citealp{Carleton2018}, albeit see \citealp{Sales2019}). 


Here we study UDG1137+16 (dw1137+16) seeking further observational constraints on the aforementioned UDG formation scenarios in a galaxy group environment. UDG1137+16 was first discovered using image enhancing techniques to search for new dwarf galaxy candidates from publicly available Sloan Digital Sky Survey (SDSS) data around the UGC 6594 group \citep{Muller2018}. Dubbed `dw1137+16',\ the authors reported that its large angular size and low surface brightness are suggestive of a good UDG candidate if located in the UGC 6594 group.  We note that, without a confirmation of this distance we cannot properly convert observed photometric properties (e.g., half-light radius) into physical units. Additionally, \citet{Muller2018} suggest that: "Better photometry is needed to derive the structural parameters more accurately", motivating the new observations presented here. 


In this work we obtain and analyse new, deep imaging for UDG1137+16 from the 0.7m Jeanne Rich Telescope (Section \ref{sec:JRdata}). We also acquire and analyse new, deep integral field spectroscopy from the Keck Cosmic Web Imager (KCWI) on the 10m Keck II telescope, deriving a recession velocity and a stellar velocity dispersion for UDG1137+16 (Section \ref{sec:IFU}). In Section \ref{sec:discussion} we discuss both our imaging and spectroscopic results in the context of UDG formation. We then supplement our measurement with those from the literature to further probe the formation of UDGs. We place a particular emphasis on the velocity dispersions of UDGs and the resulting dynamical masses inferred from them. Finally, we examine the constraining power that current dynamical masses derived from velocity dispersions have on the total halo mass of UDG halos. Concluding remarks are presented in Section \ref{sec:conc}.

\section{Imaging} \label{sec:JRdata}
\subsection{Data Acquisition}

The photometric data used in this work were acquired using the 0.7-m Jeanne Rich Telescope (JRT) and an FLI09000 CCD camera (pixel scale 1.114$\arcsec$/pix), at the Polaris Observatory Association site near Frazier Park, California. A full description of the instrumental setup is given in \citet{Rich2019}. Observations were taken in the SDSS $r$-band on the nights 2019 February 21 (16x300s), February 27 (10x300s) and March 3 (22x300s) in dark, photometric sky conditions for 4 hours of total exposure time. The seeing of the final image is $\sim$ 3.5$\arcsec$. 

Data reduction was carried out with the usual bias and dark subtractions. Images were flat-fielded using a master flat obtained via an automatic pipeline in which all the science images observed during all the dark nights of the run were masked with \texttt{NoiseChisel} \citep{Akhlaghi2015}, normalised and subsequently combined. This produces a flat with high stability and efficiency at low surface brightness levels. Images were astrometrised using the \texttt{SCAMP} software \citep{Bertin2006} and photometry was referenced to SDSS to an accuracy of $\sim \pm 0.03$ mag. The co-addition of frames was performed using a pipeline with a resistant median-based combination algorithm, efficiently rejecting satellite tails, cosmic rays and other artefacts. The depth of the final coadded image is 28.4 $\mathrm{mag\ arcsec^{-2}}$ (3 sigma in a 10x10$\arcsec$ box) following the definition provided in appendix A of \citet{Roman2020}. 

In Figure \ref{fig:overview} we display a central 5' $\times$ 5' cutout of the resulting $\sim$ 1.1 $\times$ 1.1 degree stacked image centred on the objects of interest to this work.


\begin{figure*}
    \centering
    \includegraphics[width = 0.95 \textwidth]{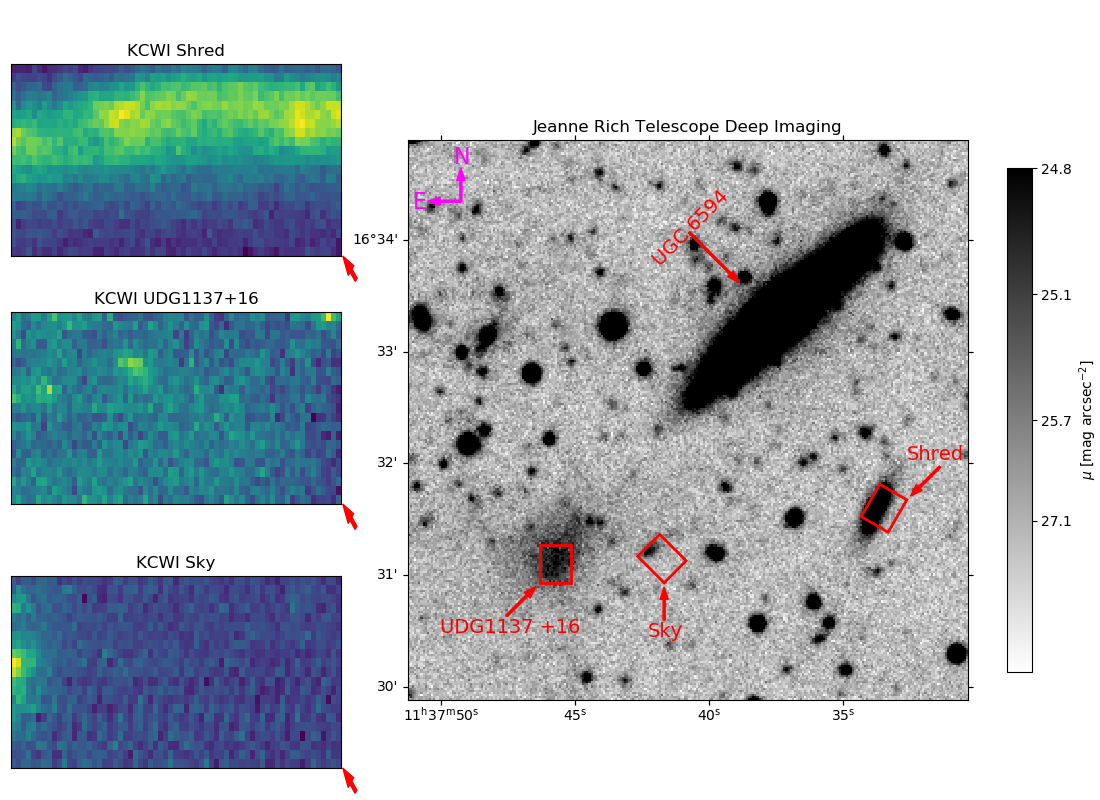}
    \caption{A visual overview of both the Jeanne Rich Telescope imaging data and the KCWI integral field spectroscopy. \textit{Left:} Whitelight depictions of the KCWI data cubes for: the object being shredded in the west of the frame (Shred, top), the UDG1137+16 (centre) and the positioning of sky exposures chosen to include a small dwarf galaxy (bottom). Red arrows on the left correspond to the same corner of the data cube as indicated by the labelling in the image on the right. Whitelight data cube image stretching is defined by the min/max pixel values of each individual cube. \textit{Right}: We display a 5' $\times$ 5' box centred on the UGC 6594 group out of the full $\sim$ 1.1 $\times$ 1.1 degree r-band Jeanne Rich Telescope mosaic. The position of the KCWI image slicers are indicated by the labelled red rectangles. North is up and east is to the left as indicated. A colour bar indicating the surface brightness of the image has been included. }
    \label{fig:overview}
\end{figure*}

\subsection{Photometric Properties}
\subsubsection{Surface brightness profile}\label{Sec2.5}
The $r$-band, major-axis surface brightness profile of UDG1137+16 was extracted from the deep JRT image following the prescription in \citet{Dullo2017}. To avoid light contamination from compact sources, background galaxies and bright foreground stars, we created an initial mask for the image with {\sc SExtractor} \citep{Sextractor} and then combined these with further manual masks as required. The IRAF task {\sc ellipse} \citep{Jedrzejewski1987} was run on the final masked image to fit elliptical isophotes to the sky-subtracted  $r$-band, JRT image of the UDG.  In extracting the light profiles, the isophote centre is held fixed, while the position angle (P.A.) and ellipticity were left as free parameters. Figure \ref{fig:photometry} shows the surface brightness, P.A. and ellipticity profiles for the galaxy. Consistency checks were performed fitting the galaxy in both 2D using \texttt{IMFIT} and in 1D with non-fixed ellipse centres when extracting the surface brightness profile. Both methods were in good agreement with the results presented herein (see further \citealp{Venhola2017} for a discussion of 1D vs 2D UDG fitting). 


\subsubsection{Light profile decomposition}\label{sec:photometry}

The \citet{Sersic1968} $R^{1/n}$ model is known to provide a good description of the underlying stellar light distributions of UDGs and other low-luminosity (\mbox{$M_{V} \ga $ -21.5 mag}) galaxies (e.g.,\ \citealt{Graham2005, Dullo2019b, Forbes2020b})\footnote{We note that some irregular morphology UDGs are known to exist where this will not be true \citep{Greco2018, Prole2019b}.} . This model is written as:

\begin{equation}
I(R) = I_{\rm e} \exp \left\{ - b_{n}\left[ 
\left(\frac{R}{R_{\rm e}}\right)^{1/n} -1 \right]\right\},
\label{Eq1}
\end{equation}

where $ I_{\rm e}$ is the intensity at the half-light radius ($R_{\rm e}$). The quantity $b_{n}\approx 2n- 1/3$ for $1\la n\la 10$ \citep{Caon1993}, is a function of the S\'ersic index $n$, and it ensures that $R_{\rm e}$ encloses half of the total luminosity.

\begin{figure}
 \includegraphics[width = 0.48 \textwidth]{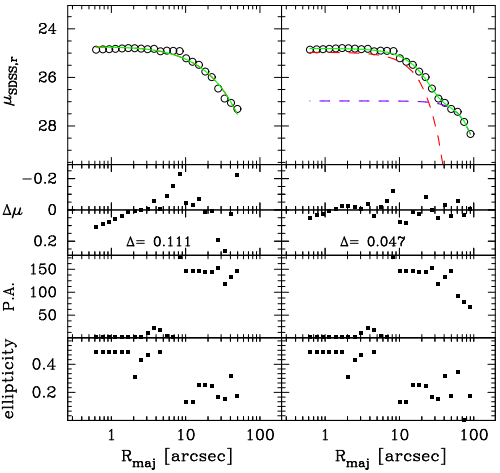}    
\caption{Major-axis surface brightness, position angle (P.A., measured in degrees from north to east) and ellipticity profiles of UDG1137+16 based on  deep imaging from the JRT (open circle). \textit{Left:} A single S\'ersic model fitted to the 1D surface brightness profile within 50$\arcsec$. \textit{Right:} The best fitting double-S\'ersic model to the full UDG light profile. We fit a central S\'ersic (red dashed curve) and an outer S\'ersic component (magenta dashed curve). The green, solid  curve is the total fit to the profiles. Our data provides mild evidence for two stellar components in UDG1137+16.}
\label{fig:photometry} 
\end{figure}

Before fitting, we calculate the FWHMs of a Gaussian point spread function (PSF) using several bright, unsaturated stars in the JRT image of the UDG. We then fit a S\'ersic model, that is convolved with the PSF in 2D, to the galaxy's 1D major-axis surface brightness profiles (e.g., \citealt{Trujillo2001, Dullo2019b, Dullo2019, Forbes2020b}) to obtain the best-fitting structural parameters that describe the galaxy 1D light profile. We do this by iteratively minimising the rms residual using the Levenberg-Marquardt optimisation algorithm. Figure~\ref{fig:photometry} shows our fits to the $r$-band light profile of UDG1137+16 along with the corresponding rms residual profiles. 



We fit our 1D light profile in two ways. Our first method is to perform a representative single S\'ersic fitting for best comparison to literature methods. We do this excluding the three outermost data points in our surface brightness profile. These points lie $>$ 50$\arcsec$ with surface brightness fainter than 27.4 mag $\mathrm{arcsec^{-2}}$ (compared to the data limit of 28.4 mag $\mathrm{arcsec^{-2}}$). Although, our single S\'ersic fit ($n \sim 1.05 \pm 0.05$ and $R_{\rm e} \sim 32.3 \pm 1.4 \arcsec$) is in good agreement with the SDSS-$r$ band data modelled by \citet{Muller2018} ($n \sim 0.89 \pm 0.1 $ and $R_{\rm e} \sim 26.6 \arcsec$), we obtain a slightly larger $R_e$. We suggest this has arisen due to the SDSS images used in \citet{Muller2018}'s analysis having insufficient depth to properly constrain the galaxy's stellar light distribution at large radii ($R \ga 15\arcsec$).

Our second method is to fit the full light profile using a two-component, double-S\'ersic model following \citet{Dullo2013}. They fitted a double S\'ersic model to the light profile of the elliptical galaxy NGC 5576 which has a long tidal tail. Our double-S\'ersic decomposition of the 1D UDG1137+16 profile gives a central UDG stellar body along with an outer stellar component, Figure \ref{fig:photometry} right. While the double-S\'ersic model gives a better description for the surface brightness profile, and hence stellar components of UDG1137+16, this fit is not designed to produce global properties of the galaxy (e.g. half-light radius) before a possible tidal interaction with UGC6594. We discuss the physical interpretation of this second stellar component in Section \ref{sec:tidal}.

As a consistency check we also perform this fitting on publicly available shallower data of UDG1137+16 from the DECaLS survey \citep{DECALS}. There is strong agreement of the extracted 1D surface brightness profiles between the JRT and DECaLS imaging within $R\sim 0.6\arcsec-20\arcsec$. The higher resolution DECaLS imaging also reveals a possible unresolved, low mass nucleus for UDG1137+16 ($L_{r} \sim 3 \times 10^{4} \mathrm{L_{\odot}}$). Additionally, analysis of the DECaLS $g$- and $r$-band data for UDG1137+16 yields a global $g-r$ colour of 0.65 $\pm$ 0.35. We measure this colour within R $<4\arcsec$ as that is where the DECaLS $g$-band data provides the best constraints. The DECaLS data have insufficient depth to properly constrain the outskirts of UDG1137+16 and so for the remainder of this paper we focus only on the JRT imaging.

  
Table~\ref{Table1} lists the best-fitting structural parameters for both the representative single and double S\'ersic fits to our JRT surface brightness profile. $1\sigma$ uncertainties are determined after running a series of Monte Carlo simulations.

We create more than 100 realisations of the UDG composite light profile using the residual profile obtained after subtracting the actual galaxy light profile from the fitted model (Figure \ref{fig:photometry}. We account for errors arising from inaccurate sky subtraction and for possible light contamination from bright foreground objects, background galaxies and faint, extended stellar halos while performing this process. In order to determine the error associated with inaccurate sky subtraction and possible light contaminations, the median background values were first measured from several 10 $\times$ 10 pixel boxes away from both the UDG and other sources in the sky subtracted image. The error is determined by the standard deviation about the mean of the median values in these boxes. These realisations were then modelled akin to the modelling of the actual galaxy light profile to derive the $1\sigma$ errors from the best-fitting parameters.

\subsubsection{UDG1137+16 Luminosities and Stellar Masses}

The total integrated luminosities ($L_{r}$) for the single- and double- S\'ersic components from our decompositions of the UDG are computed using the best-fitting major-axis structural parameters, together with the ellipticities of the individual fitted components. We assume a distance of 21.1 Mpc for the galaxy \citep{Tully2016} based on an assumption of association with the nearby UGC 6594 group (see Section \ref{sec:kinematics} for a confirmation of this assumption). All magnitudes are quoted in the AB system unless stated otherwise.

Using our adopted distance we convert the apparent magnitudes into absolute magnitudes ($M_{r}$) and then into luminosities (in solar units) assuming $M_{r}$ = +4.65 for the Sun \citep{Wilmer2018}. These magnitudes are not corrected for the small foreground Galactic extinction ($<0.1$ mag; \citealp{Schlafly2011}). We calculate a stellar-mass to light ratio ($M_{*}/L_{r}$) $\approx$ 2 $\pm$ 1.1 for UDG1137+16 using the $r$-band relation of \citet{Into2013} and our $g-r$ colour of 0.65. This yields a total stellar mass for the UDG of $1.4 \pm 0.2 \times 10^{8}\ \mathrm{M_{\odot}}$\footnote{We exclude the error in the stellar-mass to light ratio when calculating errors on the total stellar mass.}. See Table \ref{Table1} for a full summary of UDG1137+16 luminosities and stellar masses.


\begin{table*}
\setlength{\tabcolsep}{0.0620in}
\begin {minipage}{177mm}
\caption{Structural parameters of \mbox{UDG1137+16} }
\label{Table1}
\begin{tabular}{@{}llccccccccccccccccccccccccccccccccccccccccccccccc@{}}
\hline
Fit&$ \mu_{\rm e}$ & $R _{\rm e}$ &$R_{\rm e}$&$n$&$\mu_{\mathrm{0}}$&$m_{r}$&$M_{r}$&$L_{r}$& $M_{*}$\\                             
  &(mag arcsec$^{-2}$)&(arcsec)&(kpc)&&(mag arcsec$^{-2}$)&(AB mag)&(AB mag)&($\times10^{8} L_{\sun}$)&($\times 10^{8} M_{\sun}$)\\
\multicolumn{5}{c}{} \\
\hline
 Single S\'ersic   & 26.5 (0.1)   & 32.3 (1.4) & 3.3 (0.1) & 1.05 (0.05) & 24.62 (0.04) & 16.7 (0.2) & -15.0 (0.2) & 0.7 (0.1) & 1.4 (0.2)\\
\hline
 Inner S\'ersic  &  25.8 (0.1)  & 15.3 (1.0) & 1.6 (0.1)  & 0.55 (0.05) & 24.94 (0.03) & 17.8 (0.2) & -13.8 (0.2) & 0.2 (0.05) & 0.4 (0.1)\\
 Outer S\'ersic &  27.5 (0.1) & 62.3 (1.6) & 6.4 (0.2)  & 0.40 (0.11) & -- & 16.6 (0.3) & -15.0 (0.3) & 0.7 (0.2) & 1.4 (0.4) \\
 Total I+O & --  & -- & -- & -- & 24.94 (0.03) & 16.5 (0.4) & -15.2 (0.4) & 0.9 (0.5) & 1.8 (1.0)\\
\hline
\end{tabular} 

Notes.--- Structural parameters from the 1D single and double S\'ersic model fits to the $r$-band surface brightness profiles of UDG+1137. Errors in parameters are given in (brackets) following them. I+O = Inner S\'ersic + Outer S\'ersic. Stellar masses were calculated assuming M$_{\mathrm{*}}$/L $\approx$ 2 in $r$-band. We assume a distance of 21.1 Mpc (i.e., association with UGC 6594) when converting distance dependant parameters. Central surface brightnesses come from extrapolation of the fit to the centre. 
\end{minipage}
\end{table*}

\section{Integral Field Spectroscopy} \label{sec:IFU}
\subsection{Data Acquisition and Reduction}
The integral field spectroscopy used in this work was obtained on the night of 2020, February 17th. Our observations were made using KCWI on the Keck II telescope (Program ID: W140, PI: Forbes). We used the medium slicer with the BH3 grating to maximise spectral resolution ($\sigma_{\mathrm{inst}} \sim 13$ $\mathrm{km\ s^{-1}}$), enhancing our ability to recover lower velocity dispersions. We set a central wavelength of 5080 \AA\ in order to allow measurement of both the H$\mathrm{\beta}$ and Mgb triplet absorption features assuming the UDG is part of the UGC 6594 group. Offset sky exposures were observed (as depicted in Figure \ref{fig:overview}), to be used as inputs to a principal component analysis based sky subtraction, optimised for faint galaxies in the UDG regime as described in \citet{Gannon2020}. An additional pointing of KCWI with this configuration was taken on a nearby galaxy undergoing shredding (see Figure \ref{fig:overview}). Total exposure times were 16800s on the UDG, 8400s on sky and 1200s on the galaxy undergoing shredding (we dub this galaxy `Shred' for the remainder of this work).

Initial data reduction was performed using the standard KCWI pipeline \citep{Morrissey2018}. Following this, we took the output \textit{ocubes} which were non sky--subtracted and standard star calibrated, cropping them both spatially and spectrally to the regions of full coverage. We then performed an additional flat fielding step to correct for low level gradient remaining in the data (see further - \citealp{Gannon2020}). Spectra of the UDG were extracted using the KCWI data in light-bucket mode, collapsing the data cube into a single spectrum. Sky spectra were extracted in a similar manner from the offset sky exposures, avoiding the area of the data containing the dwarf galaxy. We took our extracted sky spectra and used them to perform a principal component analysis sky subtraction as described in \citet{Gannon2020}. A model for the UDG emission is required to perform this sky subtraction, we therefore selected 12 models of varying spectral type (K-type giant, A, F, G), metallicity (-1.3$\le$ [Fe/H] $\le$ -0.1) and alpha enhancement ([$\mathrm{\alpha}$/Fe] = 0 or 0.4) from the \citet{Coelho2014} library of high resolution synthetic stellar populations to fit to our data. We took the best fitting template (K-type giant; [Fe/H] = $-$1; [$\mathrm{\alpha}$/Fe] = 0) and used it as the galaxy emission model to sky subtract our spectra as part of the sky subtraction routine. Sky subtraction for both the dwarf galaxy in the sky frame and `Shred' were performed by the subtraction of on-chip sky. After applying the relevant barycentric corrections \citep{Tollerud2013}, the sky--subtracted spectra were median stacked. For our UDG spectrum we estimate a final signal to noise ratio of 20 per \AA\ in the continuum. 

\subsection{Recession Velocity, Velocity Dispersion and Dynamical Mass}
\begin{figure*}
    \centering
    \includegraphics[width = 0.95 \textwidth]{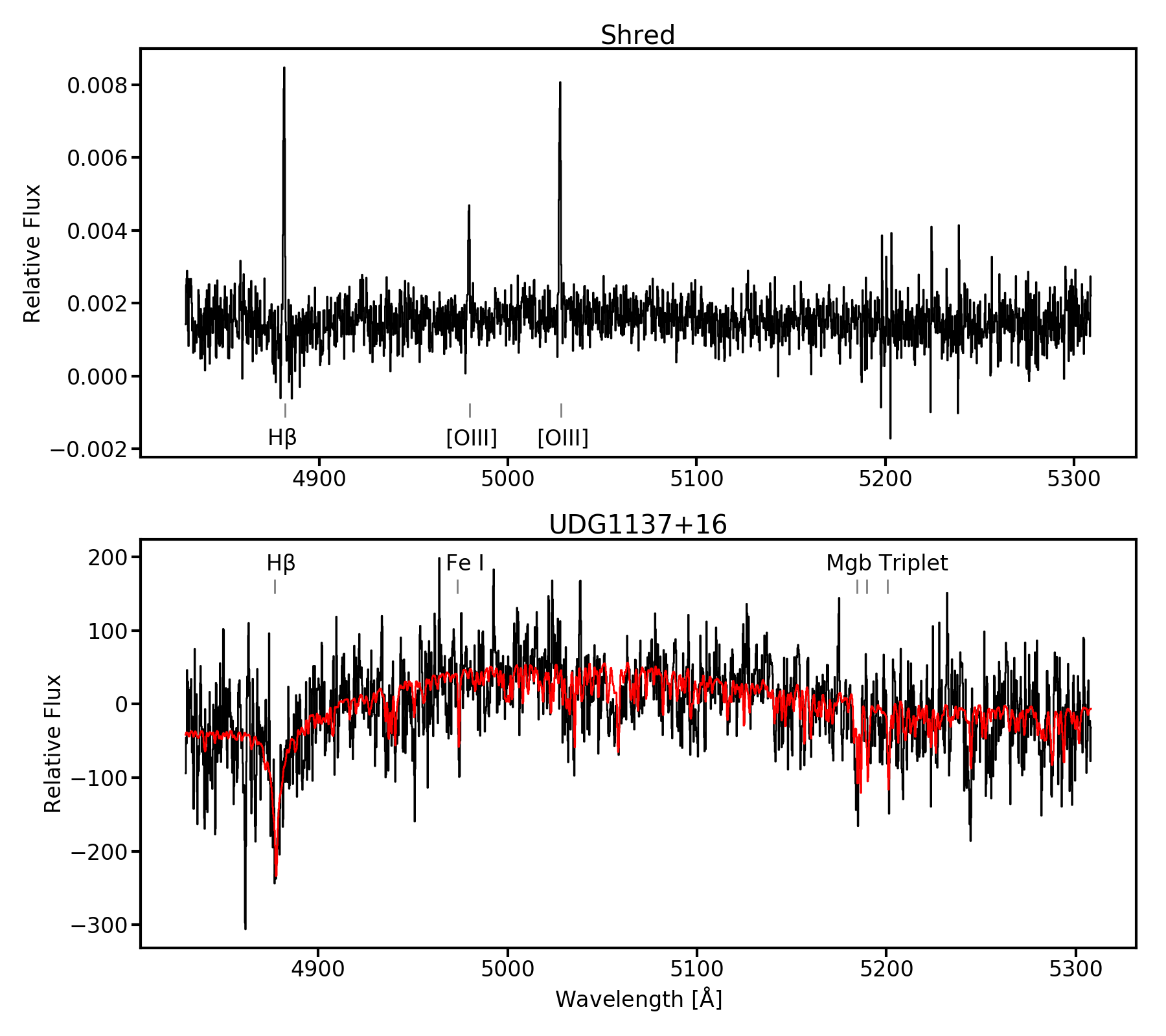}
    \caption{KCWI Spectra. \textit{Upper:} The spectrum extracted for the `Shred' object. $\mathrm{H\beta}$ and [OIII] emission lines are labelled. Based on Gaussian fits to these emission lines we measure a recession velocity of 1237 $\pm$ 3 $\mathrm{km\ s^{-1}}$ confirming its association with the low density UGC 6594 group. \textit{Lower:} The spectrum extracted for UDG1137+16 (black) and a representative \texttt{pPXF} fit (red). $\mathrm{H\beta}$, Fe I and Mgb absorption features have been labelled. Based on our fitting of the UDG1137+16 spectrum we measure a recession velocity of 1014 $\pm$ 3 $\mathrm{km\ s^{-1}}$ and a velocity dispersion of 15 $\pm$ 4 $\mathrm{km\ s^{-1}}$. }
    \label{fig:kinematics}
\end{figure*}

\subsubsection{Dwarf Galaxy in Sky Frame and `Shred'}

We display our extracted spectrum for the `Shred' galaxy in Figure \ref{fig:kinematics}. In order to measure a recession velocity from this spectrum we identify the three clear emission lines as H$\mathrm{\beta}$ and the [OIII] doublet, and fit Gaussian profiles to each to measure their centroid.  Based on the centroids of these Gaussian fits we measure a recession velocity of 1237 $\pm$ 3 $\mathrm{km\ s^{-1}}$ and thus confirm the association of this object with the low density UGC 6594 group (UGC 6594, $V_{sys} = $ 1037 $\pm$ 2 $\mathrm{km\ s^{-1}}$; \citealp{vanDriel2016}). 

We also attempt to extract a spectrum for the dwarf galaxy in the sky frame. Unfortunately, the spectrum we extract has a S/N too low to perform analysis.

\subsubsection{UDG1137+16} \label{sec:kinematics}
The extraction of a velocity dispersion in the UDG regime is known to be a particularly onerous task as UDG spectra often have a combination of low S/N, exotic chemical abundances (e.g. \citealp{Martin-Navarro2019, Ferre-Mateu2018}), and velocity dispersions at or below the instrumental resolution. Based on our previous investigation of these issues in \citet{Gannon2020}, we chose to fit our spectrum with \texttt{pPXF} \citep{Cappellari2013, Cappellari2017} using a wide ranging set of input parameters and fitting templates. For our templates we chose to use the synthesised stellar library of \citet{Coelho2014} along with a KCWI observation of the Milky Way GC M3. 

The KCWI observation of M3 was taken on 2019, April 2nd in the same KCWI configuration as our science observation but with lower central wavelengths to account for the assumed difference in redshift between our template and the target. Being observed in the same KCWI configuration it ideally models the instrumental resolution, removing the mischaracterisation of instrumental resolution as a possible source of error in our fitting. The \citet{Coelho2014} library has $>$3000 spectra with metallicities 0 < [Fe/H] < -1.3 containing both solar and alpha-enhanced mixtures at high resolution \mbox{(R = 20 000)} reducing the possibility of template mismatch in fitting the data. 

\begin{figure}
    \centering
    \includegraphics[width= 0.45 \textwidth]{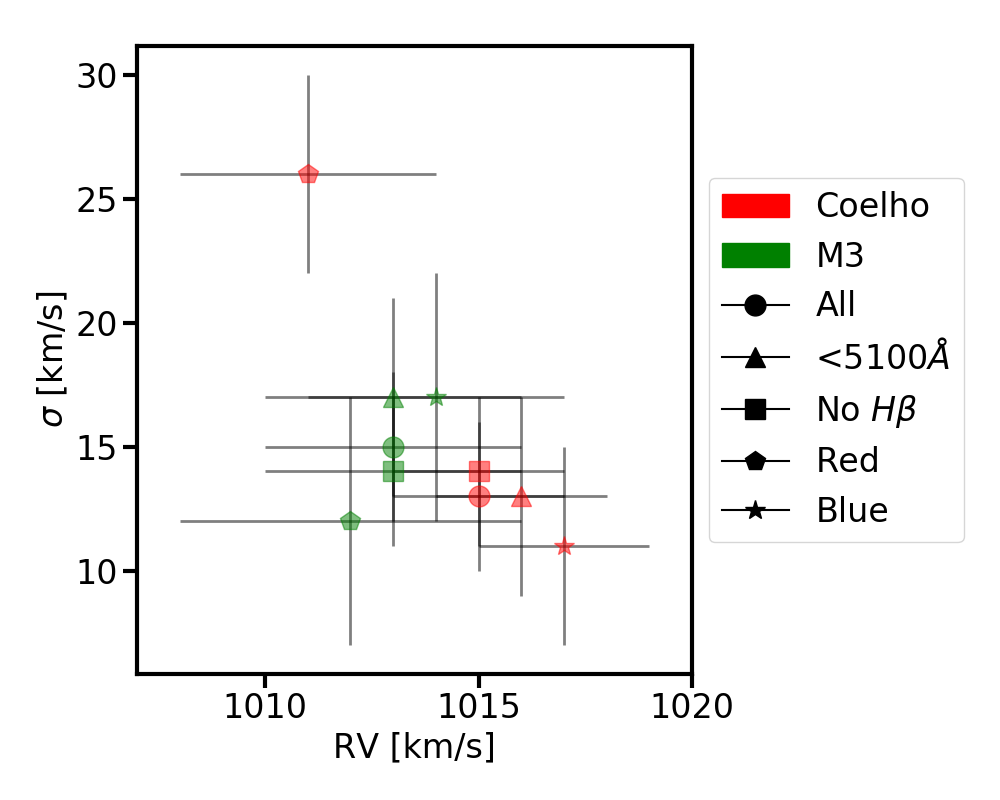}
    \caption{The recession velocity and velocity dispersion results of our \texttt{pPXF} fitting for our UDG1137+16 spectrum. Data are included for fitting with each of two template sets: the \citet{Coelho2014} synthesised stellar library (red) and an observed template from the Milky Way GC M3 (green). Fitting for five spectral regions is included for each template set: the entire spectrum (All - circular points), the spectrum less than 5100 \AA~ (<5100 \AA - triangles), the spectrum excluding the initial 175 pixels to exclude the H$\mathrm{\beta}$ sky subtraction residual (No H$\mathrm{\beta}$ - squares), the red half of the spectrum (Red - pentagons) and the blue half of the spectrum (Blue - stars). We take the average of these fits to obtain our adopted recession velocity (1014 $\pm$ 3 $\mathrm{km\ s^{-1}}$) and velocity dispersion ($\sigma$ = 15 $\pm$ 4 $\mathrm{km\ s^{-1}}$) of UDG1137+16.}
    \label{fig:ppxf_fits}
\end{figure}

We take these templates and fit our spectra with \texttt{pPXF} in 241 different input configurations. These configurations contain both pure Gaussian fitting and those with the higher Gauss-Hermite moments h$_{3}$ and h$_{4}$ also. Our \texttt{pPXF} configurations also include the addition of a wide range of both additive and multiplicative polynomials (0-10th order for both respectively) to the templates to correct for possible errors in recovery of the UDG spectrum. Moreover, we fit the spectrum in five distinct regions to test that our recovered velocity dispersion is not driven by any particular spectral region and is not adversely effected by the accidental inclusion of sky residuals in the fitting. We discard fits with reported errors in recession velocity and/or velocity dispersion that are greater than 25 $\mathrm{km\ s^{-1}}$ as we deem the fit to have ineffectively modelled our data. We display one fit, that we deem `good', in Figure \ref{fig:kinematics}. Here we fit a pure Gaussian profile with 4 additive and 4 multiplicative polynomials to the full UDG spectrum using the \citet{Coelho2014} library.

We display the median kinematic values (i.e., recession velocity and velocity dispersion with associated errors) resulting from the exhaustive fitting of our spectrum in Figure \ref{fig:ppxf_fits}. We find good agreement between fits done with both of our templates. We therefore take the average of the M3 and \citet{Coelho2014} library results as our final recession velocity = 1014 $\pm$ 3 $\mathrm{km\ s^{-1}}$ and velocity dispersion = 15 $\pm$ 4 $\mathrm{km\ s^{-1}}$. As the fitting with the M3 template is robust to a possible instrumental resolution mischaracterisation we expect this to not be a major source of error in our result. As fitting with the \citet{Coelho2014} library minimises the possibility of template mismatch, we also expect this to not be a major source of error in our result. Finally, fits to 5 different spectral regions of the spectrum report similar velocity dispersions which are consistent within errors, suggestive that a particular wavelength region does not drive our final reported values.


In the double S\'ersic decomposition described in Section \ref{sec:photometry}, the half-light radius of the central stellar body in UDG1137+16 was found to be 15.4$\arcsec$ . Due to the slightly off-centre pointing of the $\sim$ 16$\arcsec$ $\times$ 20$\arcsec$ medium slicer of KCWI, our velocity dispersion measurement represents a flux-weighted measurement within approximately one half-light radius of the central stellar body. Here the central regions are slightly over-represented in our measurement, as they are fully sampled by the slicer, while the outer regions become increasingly under-sampled in comparison. When using the single S\'ersic half-light radius, our velocity dispersion only represents a flux-weighted measurement made within $\sim$ 1/3 of the half-light radius. 



In order to convert our stellar velocity dispersion into a dynamical mass we use the mass estimator of \citet{Wolf2010}. Under the assumption of a dispersion-dominated system, this estimator has proven accurate, relying on only the 2D projected half-light radius ($R_{e}$) and the luminosity--weighted average line-of-sight velocity dispersion within this radius ($\sigma_{e}$) to determine the dynamical mass within the 3D deprojected half-light radius ($r_{1/2}$). It takes the form:

\begin{equation} \label{eqtn:wolf}
M(<r_{1/2}) = 930 (\frac{\sigma_{e}^{2}}{\mathrm{(km\ s^{-1})^{2}}}) (\frac{R_{e}}{\mathrm{pc}}) \mathrm{M_{\odot};}\quad \mathrm{where}\ r_{1/2} \approx \frac{4}{3} R_{e}
\end{equation}

We extrapolate our stellar velocity dispersion measurement out to the single S\'ersic half-light radius assuming a flat velocity profile for UDG1137+16. Using this velocity dispersion and our calculated half-light radius from the single S\'ersic fitting we infer a dynamical mass $M(<3.9$ kpc$) = 6.18 \pm 3.48 \times 10^{8} \mathrm{M_{\odot}}$. Combining this with our earlier luminosity measurement from JRT photometry $0.7 \pm 0.1 \times 10^{8} \mathrm{L_{\odot}}$) we infer an $r$-band M/L ratio of $17.7 \pm 10.2$ within the same radius.


\section{Discussion} \label{sec:discussion}

Due to their extremely similar recession velocities we suggest UDG1137+16 is likely part of the low density UGC 6594 group (UGC 6594, $V_{sys} =$ 1037 $\pm$ 2 $\mathrm{km\ s^{-1}}$; UDG1137+16, $V_{sys} =$ 1014 $\pm$ 3 $\mathrm{km\ s^{-1}}$) placing it at a distance of 21.1 Mpc. Based on the JRT data and this assumed distance we note UDG1137+16 meets the standard definition for a UDG ($\mu_{0,g} > 24\ \mathrm{mag\ arcsec^{-2}}$; $R_{e} > 1.5$kpc), irrespective of the method we use to fit our JRT surface brightness profile. We therefore confirm \citet{Muller2018}'s hypothesis that this galaxy is a UDG. We summarise the properties of UDG1137+16 in Table \ref{tab:113716}.

\subsection{Extended Stellar Features} \label{sec:tidal}

\begin{figure}
    \centering
    \includegraphics[width = 0.45 \textwidth]{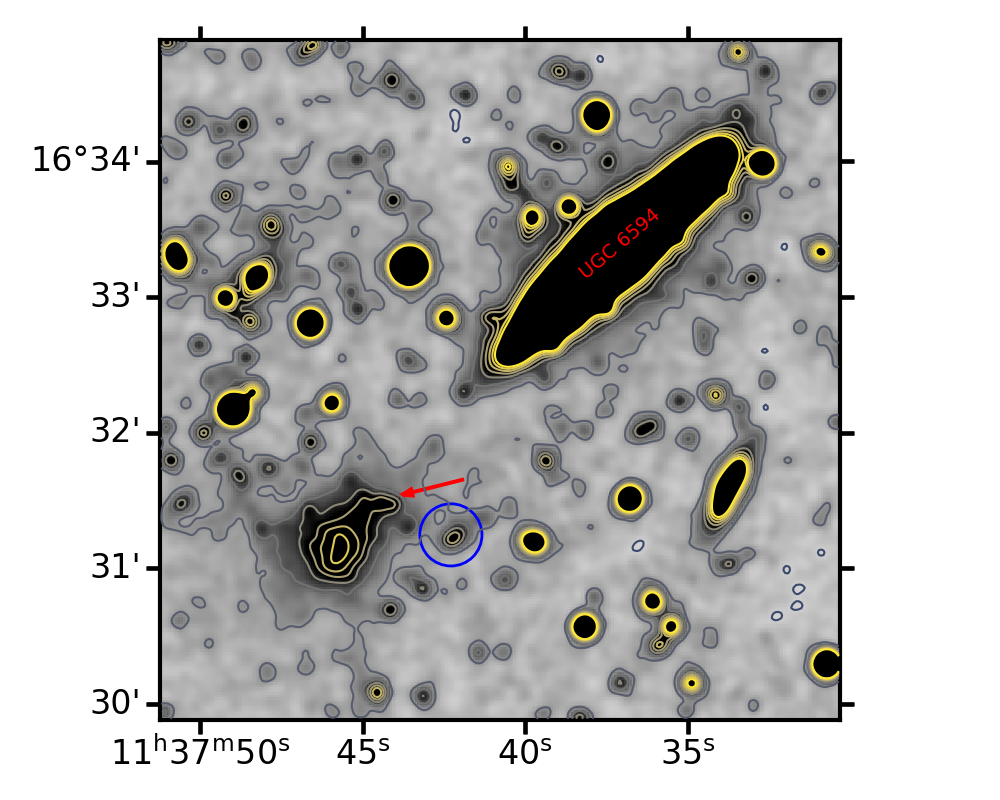}
    \caption{A smoothed, contoured version of the imaging data displayed in Figure \ref{fig:overview}. We smooth the 5' $\times$ 5' JRT cutout using a 2D Gaussian kernel of width 2 pixels and apply 10 contours to help highlight low surface brightness structure in the frame. The elongation in contours directly south of UDG1137+16 is caused by a nearby bright star. A blue circle indicates the position of the dwarf galaxy in the frame which appears faintly connected to our UDG. A red arrow indicates an elongation in the UDG towards UGC 6594 suggestive of possible tidal interaction. Qualitative evidence exists for UDG1137+16 undergoing a tidal interaction.}
    \label{fig:tidal}
\end{figure}

\begin{table}
    \centering
    \begin{tabular}{ccc}
    \hline
         Property & Value & Source \\
         \hline
         R.A. [J2000] & $11:37:46$ & \citet{Muller2018} \\
         Dec. [J2000] & $+15:31:09$ & \citet{Muller2018} \\
         Dist. [Mpc] & 21.1 & Adopted \citep{Tully2016} \\
         $m_r$ [mag] & 16.7 $\pm$ 0.2 & Sec.\ref{sec:photometry}\\
         $M_{r}$ [mag] & -15.0 $\pm$ 0.2 & Sec.\ref{sec:photometry} \\
         $\mu_{0}$ [$\mathrm{mag\ arcsec^{-2}}$] & 24.62 $\pm$ 0.04 & Sec.\ref{sec:photometry} \\
         $R_{e}$ [arcsec] & 32.3 $\pm$ 1.4 & Sec.\ref{sec:photometry} \\
         $R_{e}$ [kpc] & 3.3 $\pm$ 0.1 & Sec.\ref{sec:photometry}\\
         $M_{*}$ [$\mathrm{\times 10^{8}\ M_{\odot}}$] & 1.4 $\pm$ 0.2 & Sec.\ref{sec:photometry}\\
         b/a & 0.8 & Sec.\ref{sec:photometry} \\
         RV [$\mathrm{km\ s^{-1}}$] & 1014 $\pm$ 3 & Sec.\ref{sec:kinematics}\\
         $\sigma$ [$\mathrm{km\ s^{-1}}$] & 15 $\pm$ 4 & Sec.\ref{sec:kinematics}\\
        $M_{\mathrm{dyn}}$ [$\mathrm{\times 10^{8} M_{\odot}}$] & 6.18 $\pm$ 3.48 & Sec.\ref{sec:kinematics} \\
        $M/L$ & 17.7 $\pm$ 10.2 & Sec.\ref{sec:kinematics}\\
        \hline
    \end{tabular}
    \caption{A summary of the basic properties of UDG1137+16. For basic photometric properties we use the single S\'ersic fit in Section \ref{sec:photometry} which allows the best comparison to the methods used in other works. For distance dependent quantities we use our adopted distance of 21.1 Mpc.}
    \label{tab:113716}
\end{table}

In Section \ref{sec:photometry}, the proper decomposition of the 1D surface brightness profile of UDG1137+16 required (at least) two distinct components. In order to investigate this further, in Figure \ref{fig:tidal} we display a smoothed, contoured 5' $\times$ 5' cutout of our JRT data around UDG1137+16. Visual inspection of the image reveals a clear distortion of the stellar envelope in the direction of UGC 6594 (labelled with a red arrow), indicative of a tidal interaction between the galaxies. Moreover, the extraction of our 1D surface brightness profile reveals a clear twisting of the elliptical isophotes to align with the direction of UGC 6594, a known effect of tidal interactions \citep{Carleton2018}.

We argue the close projected distance between the galaxies ($\sim 20$ kpc between galaxy centres at the adopted distance of 21.1 Mpc) and their extremely similar recession velocities provides further evidence that they are likely an interacting pair. While we do not know the precise physical separation between galaxies, if the average tangential distance along line-of-sight is similar to the projected distance, 3D separation will be on average $\sqrt{2}$ times projected 2D separation. Assuming a rough mass for UGC6594 of $10^{12}\ \mathrm{M_{\odot}}$ we calculate tidal radii for our measured dynamical mass using equation 1 of \citet{Mowla2017}. They are 1.2/2.4/3.5 kpc based on physical separations of 20/40/60 kpc with UGC6594. All these radii are less than that which our dynamical mass is measured, suggesting tidal features should form. We therefore propose a scenario where the UDG outer stellar material is redistributed via tidal interaction into a stellar envelope of approximately similar stellar mass to the central stellar body.

Our deep image does not reveal a tidal bridge between the galaxies, as may be expected for interacting galaxies. This suggests either that we have insufficient depth to detect such a feature or that this interaction is in its infancy and such structures are yet to form. It should be noted that UDGs embedded in tidal structures below the surface brightness limit of our data are known to exist (e.g. CenA-MM-Dw3; \citealp{Crnojevic2016}). Additionally, while there is no obvious evidence for widespread tidal disruption of UDGs in the Coma cluster \citep{Mowla2017}, some UDGs are known to exhibit tidal features (e.g. \citealp{Mihos2015}). 

There may be cause for concern that tidal stripping could adversely affect our stellar velocity dispersion and hence dynamical mass measurement (see e.g. \citealp{Montes2020}). However, tidal stripping operates outside--in, affecting the outskirts of the galaxy and leaving the central velocity dispersion approximately constant \citep{Bender1992, Chilingarian2009, Blom2014, Penny2015}. For example, an extreme Local Group UDG analogue with suspected tidal origins, Andromeda XIX, exhibits a largely flat velocity dispersion profile \citep{Collins2020}. Additionally, our stellar velocity dispersion measurement is made within the half-light radius of the central stellar component, which would be subject to lessened tidal forces in comparison to the outskirts of the galaxy. We conclude that tidal stripping is not likely to strongly bias our centralised velocity dispersion measurement.



The galaxy we label `Shred' in Figure \ref{fig:overview} is also located close in both projection and velocity space to UGC 6594. It shows the galaxy is being disrupted and exhibits an irregular morphology. In Figure \ref{fig:tidal} this galaxy also hints at the `S-shaped' features indicative of a tidal disruption, however the coincidental association of other nearby objects makes it difficult to ascertain whether or not these features are real or merely incidental. Additionally, our spectrum for this object displays strong emission features indicative of active star formation which is known to be enhanced by tidal interactions \citep{Martig2008}. We suggest it is likely that this galaxy is also undergoing tidal interactions in the group.

\begin{figure*}
    \centering
    \includegraphics[width = 0.95 \textwidth]{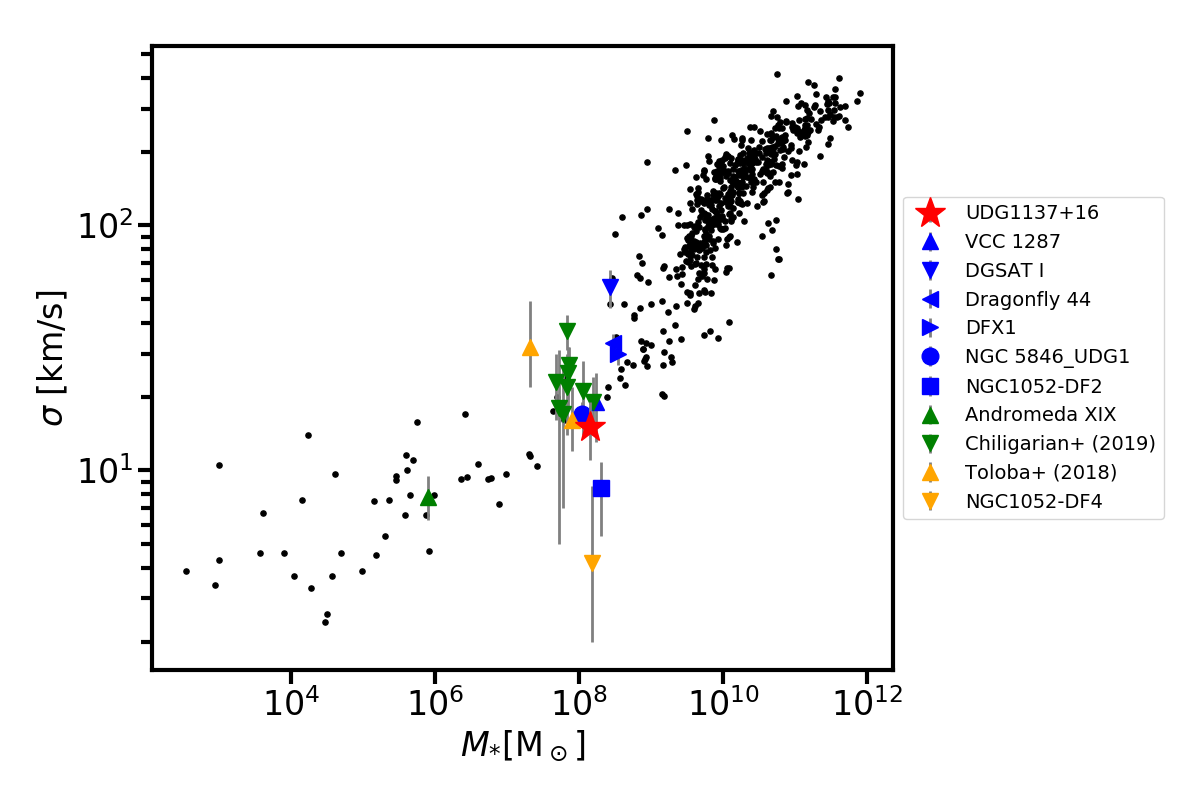}
    \caption{Central stellar velocity dispersion vs galaxy stellar mass. We plot a sample of non-UDGs: early type galaxies \citep{Cappellari2013}, elliptical/S0/irregular/spiral galaxies \citep{Harris2013}, dwarf ellipticals \citep{Chilingarian2009} and Local Group dwarfs \citep{Mcconnachie2012} (all black points). We also plot our data (red star) along with other UDGs with central stellar velocity dispersions (blue points - \citealp{vanDokkum2017, vanDokkum2018, Danieli2019, vanDokkum2019b, Martin-Navarro2019, Gannon2020, Forbes2021}) and `UDGs' from the literature where some debate remains as to their precise classification/properties (green triangles - \citealp{Chilingarian2019, Collins2020}). We also include measured GC velocity dispersions of UDGs assuming they are comparable to the stellar velocity dispersion (yellow triangles - \citealp{Toloba2018, vanDokkum2019}). For a full summary of the literature UDGs see Appendix \ref{app:summary}. Most UDGs, to date, have central stellar velocity dispersions consistent with non-UDGs for their stellar mass.}
    \label{fig:sigma_mstar}
\end{figure*}


\subsection{Formation Scenarios}

\begin{figure*}
    \centering
    \includegraphics[width = 0.95\textwidth]{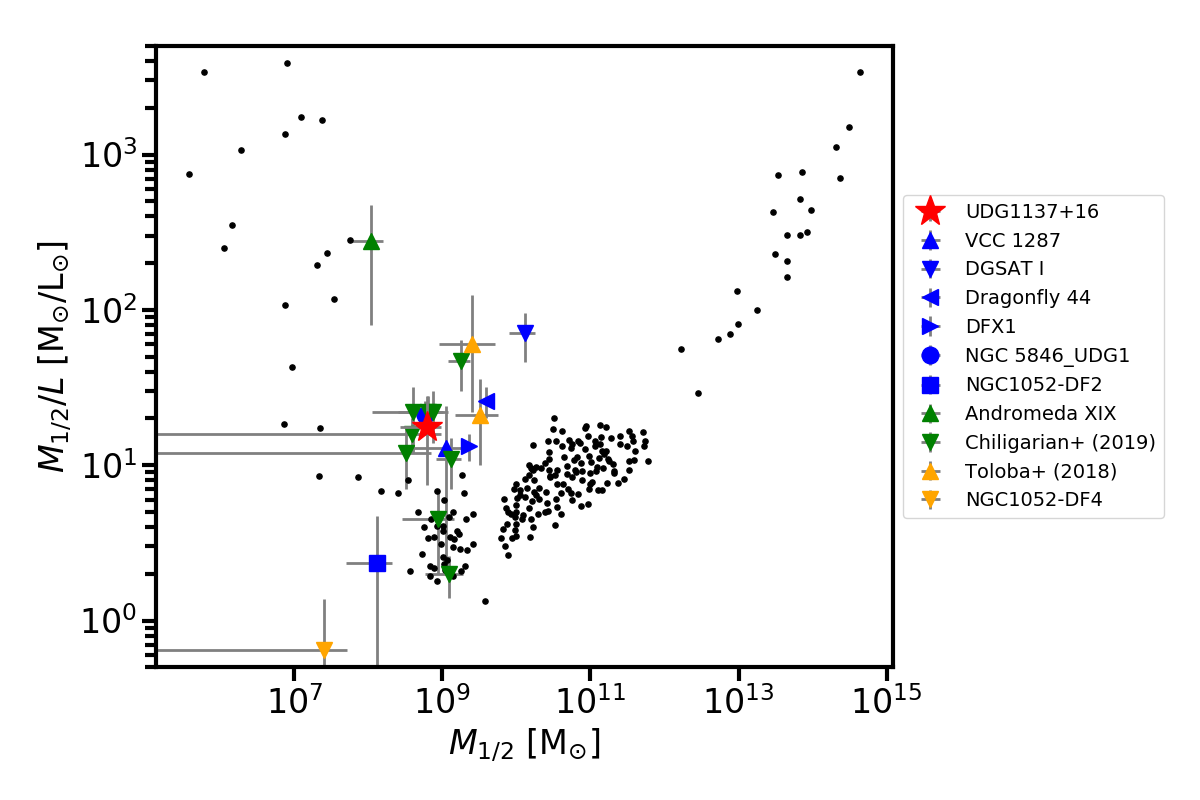}
    \caption{Dynamical mass to light ratio vs dynamical mass within the half-light radius. Non-UDGs (black points) are plotted to establish the `U-shaped' relation \citep{ Zaritsky2006, Wolf2010, Forbes2011, Cappellari2013, Toloba2014}. For UDGs, the symbols and colours are as in Figure \ref{fig:sigma_mstar}. Most UDGs tend to deviate from `U-shaped' relation. }
    \label{fig:ml}
\end{figure*}

In Figure \ref{fig:sigma_mstar} we plot our stellar velocity dispersion for UDG1137+16, along with measurements for other UDGs (\citealp{vanDokkum2017, vanDokkum2019b, Martin-Navarro2019, Danieli2019, Gannon2020, Forbes2021}) and objects called UDGs where some debate remains as to their precise classification/properties \citep{Chilingarian2019, Collins2020}. Additional to this, we include UDGs from \citet{Toloba2018}\footnote{We exclude VLSB-B in plotting the \citet{Toloba2018} sample due to its large errors.} and \citet{vanDokkum2019} with GC kinematics under the assumption their measured GC kinematics are representative of the stellar velocity dispersion of their associated UDG (see further \citealp{Forbes2021}). Our UDG sample is broadly selected with no criteria related to galaxy environment. We study the stellar velocity dispersion -- stellar mass relation using a sample of non-UDGs (dwarf ellipticals; \citealp{Chilingarian2009}, Local Group dwarfs; \citealp{Mcconnachie2012}, elliptical/S0/irregular/spiral galaxies; \citealp{Harris2013}  and early type galaxies; \citealp{Cappellari2013}). For more details on the UDG sample, including environmental associations, see Appendix \ref{app:summary}. From Figure \ref{fig:sigma_mstar} it is clear that, with the possible exception of NGC1052-DF2 and NGC1052-DF4, all UDGs appear to follow the same central velocity dispersion -- stellar mass relation.

Interestingly, there appears to be a lack of non-UDG objects at a similar velocity dispersion but higher stellar mass than UDGs (Figure \ref{fig:sigma_mstar}). Tidal stripping has been a formation mechanism \mbox{proposed} for UDGs \citep{Yozin2015, Carleton2018}. As previously discussed it is not necessarily expected to change the central velocity dispersion in a galaxy but will remove stellar mass. In a strong stripping formation scenario (e.g. $>$ 99\% mass loss; \citealp{Carleton2018}) we might expect to see progenitor galaxies for UDGs in this region of parameter space. The lack of UDG progenitor candidates at a similar velocity dispersion and a stellar mass of $\sim$ 10$^{10}$ M$_{\odot}$ disfavours strong tidal stripping as a natural formation pathway for UDGs without a mechanism to decrease the central stellar velocity dispersion.


Multiple pericentric passages my be an example of such a mechanism, although it would be expected to more significantly disrupt the morphology of the observed galaxy \citep{Errani2015}. We also note recent work demonstrating certain radial orbits can lead to both a dark matter and a stellar velocity dispersion reduction \citep{Maccio2020} in UDGs. In this case we would not expect our galaxy to follow the stellar velocity dispersion -- stellar mass relation, similar to NGC1052-DF2/NGC1052-DF4. UDG1137+16 follows the relation however, and does not appear to have lost dark matter in its central regions suggesting it is not in such a radial orbit. We suggest future work studying the resolved velocity profiles of field and group UDGs may be vital to properly understand the effects of tidal interactions on their formation and evolution.

In the case of UDG1137+16, the UDG formation simulations of \citet{Sales2019} in denser environments would suggest that it may not be formed through strong tidal interaction as it obeys the stellar velocity dispersion -- stellar mass relation. We also note that, while the excess light from the stellar envelope at large radii increases the fitted half-light radius in the single component decomposition of the surface brightness profile, the central stellar body in the multi-component fitting still meets the standard UDG definition. This suggests that UDG1137+16 was already of large size before it began its current tidal interaction. We therefore disfavour formation scenarios for UDG1137+16 that rely solely on strong tidal stripping/interactions to produce the large sizes observed in UDGs. In order to pose definitive constraints on the formation process of UDG1137+16 and other UDGs, the need of high quality spectra to perform stellar population analysis becomes imperative and will be subject of future work.

We take our calculated dynamical mass and M/L ratio and compare them to the `U-shaped' relation of non-UDGs in Figure \ref{fig:ml}. We plot a selection of non-UDGs to establish the relation \citep{Zaritsky2006, Wolf2010, Forbes2011, Cappellari2013, Toloba2014} and include those UDG objects previously plotted in Figure \ref{fig:sigma_mstar}. As has been previously established \citep{Toloba2018, vanDokkum2019b, Gannon2020}, UDGs in Figure \ref{fig:ml} do not necessarily follow the `U-shaped' relation, with some lying above and others lying below their expected M/L ratio given their dynamical mass. We find this also to be true for UDG1137+16. It lies above the standard `U-shaped' relation with a higher M/L ratio for its dynamical mass than other non-UDGs. Similar results have been found for other UDGs that are known to be associated with large GC systems (e.g. VCC 1287). Previously, the correlation between GC system richness and total dark matter halo mass (e.g. \citealp{Burkert2019}) has been used to infer two types of UDG, those with dwarf galaxy-like and those with more massive dark matter halos \citep{Forbes2020}. Unfortunately, our JRT data have insufficient resolution to adequately probe the GC system of this UDG. We can therefore only suggest that UDG1137+16 displays inner dark matter halo properties similar to other GC rich UDGs. We investigate further the total halo mass of UDG1137+16 in Section \ref{sec:DMAHMP}.


Given that only NGC1052-DF4 clearly lies below the trend in Figure \ref{fig:ml}, we focus on those UDGs lying above the `U-shaped' relation to higher M/L ratio given their dynamical mass. Noting that all the UDGs plotted are of broadly similar luminosity and stellar mass, the observed deviations from the `U-shaped' relation must be driven by their dynamical masses. Additionally, these dynamical masses are proportional to the square of the velocity dispersion and the half-light radius of the galaxy  (e.g., Equation \ref{eqtn:wolf}). We have already demonstrated in Figure \ref{fig:sigma_mstar} that UDGs do not typically exhibit abnormal stellar velocity dispersions for their stellar mass. Thus, the driver for dark matter dominated UDGs being offset towards a higher M/L ratio given their dynamical mass is, at least partially, their larger sizes compared to that of non-UDGs with similar dynamical mass lying on the `U-shaped' relation. As noted by \citet{vanDokkum2019b}: \textit{"The effective radius always contains 50 \% of the light, but it does not contain a fixed fraction of the dark matter"}. If we assume a similar dark matter halo structure, for both UDGs and the non-UDGs, then the larger half-light radii of UDGs encapsulates more of the dark matter halo. The natural effect of this is a higher dynamical mass and hence a higher M/L ratio. However, we cannot rule out the possibility that there is another parameter, so far unaccounted for, causing these deviations. 

\subsection{Dynamical Masses as Halo Mass Predictors} \label{sec:DMAHMP}
\begin{figure*}
    \centering
    \includegraphics[width = 0.95 \textwidth]{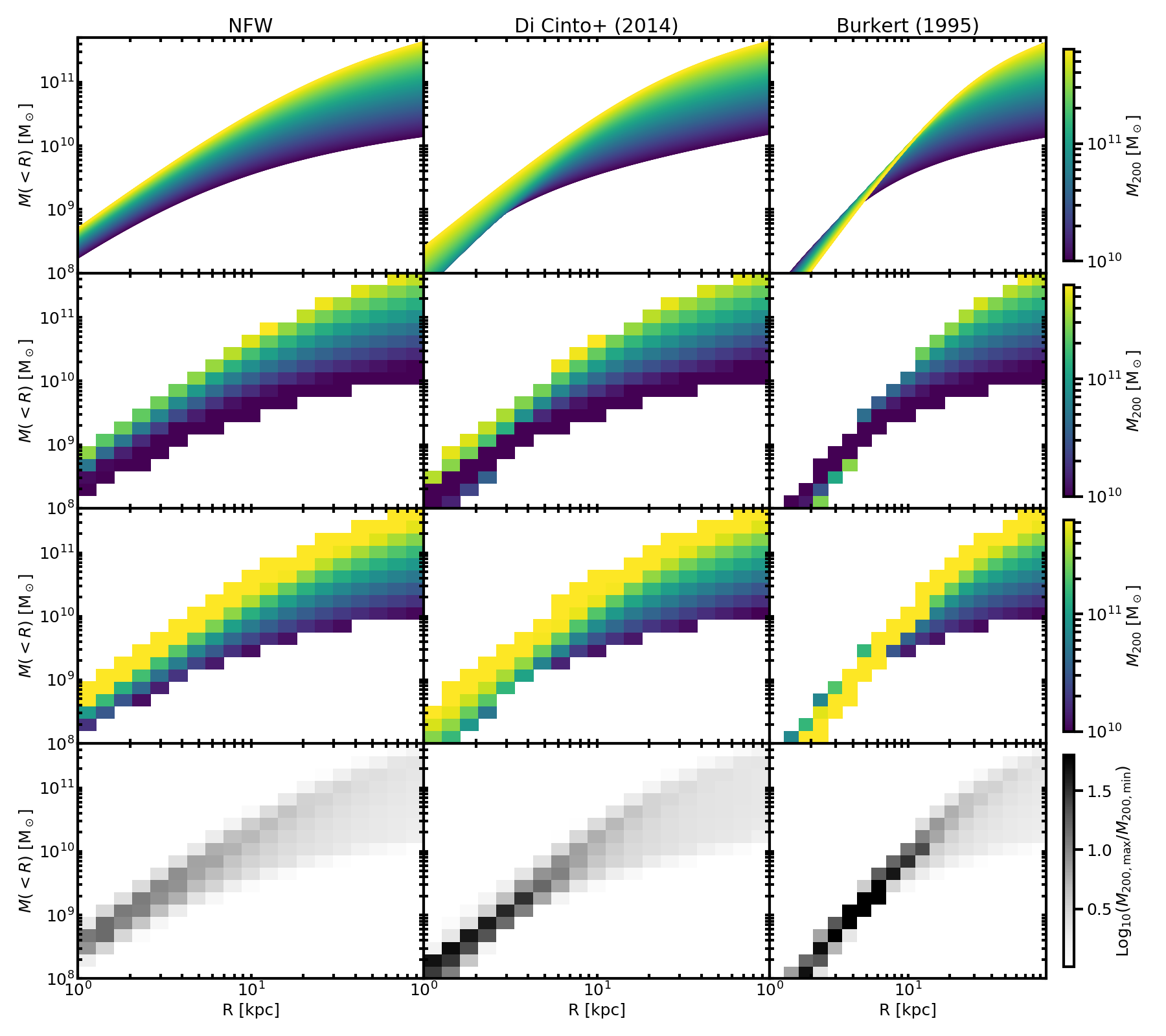}
    \caption{An investigation of the predictive power of observational mass measurements. Throughout the plot we are showing the mass enclosed within a radius from halo centre, in solar masses, ($M(<R)$) vs that radius from halo centre, in kpc, (R). We plot for three different halo profiles, that of \citet{Navarro1996} (\textit{left column}), \citet{DiCintio2014} (\textit{middle column}) and \citet{Burkert1995} (\textit{right column}). \textit{Top row:} The enclosed mass as a function of radius for each halo profile. We plot for 1000 halos with masses ($M_{200}$) equally spaced logarithmically between $10^{10}\ \mathrm{M_{\odot}}$ and the 95\% confidence upper limit of UDG halo masses ($10^{11.8}\ \mathrm{M_{\odot}}$; \citealp{Sifon2018}). \textit{Second row:} We bin the data in both radius and enclosed mass into 20 equal bins in log space for each respectively. In this row we display the minimum halo mass passing through each region. \textit{Third row:} The same as the second row, but instead plotting the maximum halo mass passing through each region rather than the minimum. \textit{Bottom row:} The logarithmic difference between the maximum and minimum halo mass in each binned region. Darker regions indicate those where measured masses poorly constrain the total halo mass of the galaxy (refer to scale bar on right). Many UDG dynamical masses fall in the $\sim$ 2-5 kpc range where they cannot effectively infer the total halo mass of the UDG, even after an assumption of halo profile. In order to better constrain the total halo mass of UDGs mass measurements at greater radii are required.}
    \label{fig:m200s}
\end{figure*}


The extrapolation of a single dynamical mass measurement into a total halo mass requires the assumption of a halo profile. Here, we use the standard cuspy \citet{Navarro1996} (NFW) halo profile, along with cored halo profiles motivated by dwarf galaxy observations \citep{Burkert1995} and simulations \citep{DiCintio2014}. We note there, exists evidence favouring cored halo profiles for UDGs from both observations \citep{vanDokkum2019b, Wasserman2019, Gannon2020} and galaxy formation simulations \citep{DiCintio2017, Carleton2018, Martin2019}. NFW and \citet{DiCintio2014} profiles are plotted using the methodology in the appendix of \citet{DiCintio2014}. For our over-density criterion we use the standard $\Delta_{200}$ where our halo mass is measured when the halo density reaches 200 times the critical density of the Universe. As the \citet{DiCintio2014} profiles also require a stellar mass for the galaxy we use a stellar mass of $M_{*} = 10^{8}\ \mathrm{M_\odot}$ which is approximately the stellar mass of the currently studied UDGs.

\citet{Burkert1995} mass profiles are calculated in a similar manner. Noting that, by definition the mass at $R_{200}$ is $M_{200}$, it is possible to recover the core radius of a \citet{Burkert1995} halo using equations 2 and 4 of \citet{Salucci2000}. Once this parameter is known equation 2 of \citet{Salucci2000} provides the mass for a \citet{Burkert1995} halo as a function of radius.

In Figure \ref{fig:m200s} we display the resulting halo mass profiles for 1000 halos with masses ($M_{200}$) equally spaced logarithmically between $10^{10}\ \mathrm{M_{\odot}}$ and the 95\% confidence upper limit of UDG halo masses ($10^{11.8}\ \mathrm{M_{\odot}}$; \citealp{Sifon2018}). We also display the minimum and maximum halo masses in each of 400 bins (20 in radius and 20 in enclosed mass). Finally, we include the logarithmic difference between the minimum and maximum halo mass in each bin in order to quantify the predictive power of differing regions of parameter space. Halos with total mass below $10^{10}\ \mathrm{M_{\odot}}$ are not plotted as both the \citet{DiCintio2014} and \citet{Burkert1995} halo profiles quickly run into numerical issues in this regime. Noting this, Figure \ref{fig:m200s} is strictly only valid for differentiating between UDGs in dwarf-like and more massive halos. It does not display the predictive power of observational measurements to distinguish between different dwarf-like dark matter halos.

For both the cored profiles of \citet{DiCintio2014} and \citet{Burkert1995}, the presence of a dark matter core leads to a complex interpretation of masses measured within $\sim$ 10 kpc. In both cases the presence of \textit{more massive} dark matter halos can lead to \textit{less} mass within the central regions as these halos become more efficient at making a core. A problematic result of this effect is that observational mass measurements coming within these central regions poorly constrain the total halo mass of the galaxy. This is not true for the cuspy NFW halo profile where, at all radii, an increase in mass corresponds to an increase in total halo mass.

In the case of our UDG1137+16 dynamical mass measurement, the best fitting NFW profile corresponds to a $\sim 2 \times 10^{9}\ \mathrm{M_{\odot}}$ halo with 1-$\sigma$ range extending from approximately $6 \times 10^{8}\ \mathrm{M_{\odot}}$ to $4 \times 10^{9}\ \mathrm{M_{\odot}}$.\footnote{Note that halos of this mass range are not plotted in Figure \ref{fig:m200s}.} At face value this would suggest an under-massive dark matter halo for UDG1137+16. The only UDG with a resolved stellar velocity dispersion profile and total halo mass fitting, Dragonfly 44 \citep{vanDokkum2019b, Wasserman2019}, requires a strong tangential anisotropy to fit a cuspy NFW profile to its data. Additionally, there is observational evidence against using NFW profiles to fit UDG data, coming from the comparison of dynamical mass measurements to total halo mass estimates from GC counts (\citealp{Gannon2020,Forbes2021}). Noting the previous observational evidence against NFW profiles for UDGs that are similarly dark matter dominated and the improbably low inferred total halo masses, we suggest it likely the cuspy NFW profile poorly describes the true halo profile of UDG1137+16. 

In the case of UDG1137+16 all \citet{DiCintio2014} halos up to a total halo mass of $\sim 4 \times 10^{9}\ \mathrm{M_{\odot}}$ lie within 1-$\sigma$ of our dynamical mass measurement. For the \citet{Burkert1995} halo profile total halo masses up to $\sim 6 \times 10^{9}\ \mathrm{M_{\odot}}$ lie within 1-$\sigma$ of our dynamical mass measurement. Once halos reach a total dark matter mass of  $\sim 10^{12}\ \mathrm{M_{\odot}}$ the \citet{Burkert1995} halos again lie within 1-$\sigma$ of our mass measurement as the halos develop a sufficiently large core to enclose masses \textit{less} than other less massive halos. We note that the \citet{Burkert1995} profiles were observationally motivated, being based on dwarf galaxies, and so may not scale well into this more massive regime. 

We conclude that it is difficult for our dynamical mass measurement to precisely recover a total halo mass for UDG1137+16. This problem will be one common to most UDGs with singular mass measurements at small radii such as those usually calculated using velocity dispersions and Equation \ref{eqtn:wolf}. Differentiation between different total halo masses requires either a mass measurement at a radius larger than what is typically available in the low surface brightness regime (e.g., from X-rays or lensing) or a resolved mass profile for the UDG (e.g., \citealp{vanDokkum2019b, Wasserman2019}). As a consequence it is entirely possible that, despite many UDGs being measured to have dark matter halos of similar inner mass, they reside in dark matter halos of quite different total mass.


\section{Conclusions} \label{sec:conc}

In this work we have studied UDG1137+16 using deep $r$-band imaging from the Jeanne Rich Telescope and deep spectroscopic data from the Keck Cosmic Web Imager. Our main conclusions are as follows:

\begin{itemize}
    \item Through analysis of the photometric properties of UDG1137+16 and our newly measured recession velocity we determine this galaxy is a bona fide UDG ($R_{e} = 3.3 \pm 0.1$ kpc; $\mu_{0,r} = 24.62 \pm 0.04\ \mathrm{mag\ arcsec^{-2}}$), confirming \citet{Muller2018}'s hypothesis. Furthermore we place it, along with a nearby galaxy undergoing a strong tidal interaction, in the low density UGC 6594 group.
    
    \item Our analysis also provides qualitative evidence for a dual stellar component in UDG1137+16. We suggest it is likely that UDG1137+16 has a stellar envelope ($M_{*} \approx 1.4 \pm 0.4 \times 10^{8} \mathrm{M_{\odot}}$), surrounding its central stellar component ($M_{*} \approx 0.4 \pm 0.1 \times 10^{8} \mathrm{M_{\odot}}$), comprising material redistributed via tidal interaction. 
    
    \item We measure a stellar velocity dispersion of $15 \pm 4\ \mathrm{km\ s^{-1}}$ and calculate a dynamical mass and mass to light ratio for UDG1137+16 within 3.9 kpc. Our measurements are consistent with the central stellar velocity dispersion -- stellar mass relation but inconsistent with the dynamical mass -- mass to light ratio relation established for non-UDGs. UDG1137+16 lies at higher mass to light ratio than non-UDGs at similar dynamical mass.
    
  
    \item We examine the relationship between central stellar velocity dispersion and stellar mass for a sample of UDGs finding that most are fully consistent with the relationship established for a sample of non-UDG galaxies. Additionally, we note that there exist few non-UDG galaxies, at similar velocity dispersions but much higher stellar masses. This suggests a lack of progenitors for strong tidal stripping formation scenarios (e.g., $>$ 99\% mass loss) which would disfavour their ability to reproduce the known UDG population. Alternatively, a tidal stripping mechanism that decreases the central stellar velocity dispersion without significantly disrupting the galaxy's morphology is needed.  
    

    \item Finally, we investigate the ability of current dynamical mass measurements based on a stellar velocity dispersion to predict total halo masses for UDGs. We find that even under reasonable assumptions for a halo profile, current lone dynamical masses based on a stellar velocity dispersion are unable to constrain total halo mass. In order to properly constrain total halo mass, mass measurements made at larger radii or a resolved mass profile are required (e.g., Dragonfly 44 \citealp{vanDokkum2019b, Wasserman2019}. This also implies UDGs that display similar dark matter characteristics at small radii may not necessarily have the same total halo masses. 
 
\end{itemize}

\section*{Acknowledgements}
We thank the anonymous referee for their careful consideration of our work and for providing useful comments to improve upon it. We thank R. Turner and A. Romanowsky for insightful conversations throughout the creation of this work. Some of the data presented herein were obtained at the W. M. Keck Observatory, which is operated as a scientific partnership among the California Institute of Technology, the University of California and the National Aeronautics and Space Administration. The Observatory was made possible by the generous financial support of the W. M. Keck Foundation. The authors wish to recognise and acknowledge the very significant cultural role and reverence that the summit of Maunakea has always had within the indigenous Hawaiian community.  We are most fortunate to have the opportunity to conduct observations from this mountain. JSG acknowledges financial support received through a Swinburne University Postgraduate Research Award throughout the creation of this work. DAF thanks the ARC for financial support via DP160101608. AFM has received financial support through the Post-doctoral Junior Leader Fellowship Programme from ``La Caixa'' Banking Foundation (LCF/BQ/LI18/11630007). JPB gratefully acknowledges support from National Science foundation grants AST- 1518294 and AST-1616598. B.T.D acknowledges supports from a Spanish postdoctoral fellowship `Ayudas 1265 para la atracci\'on del talento investigador. Modalidad 2: j\'ovenes investigadores.' funded by Comunidad de Madrid under grant number 2016-T2/TIC-2039 and from the grant `High-resolution, multiband analysis of galaxy centers (HiMAGC)' with reference number PR65/19-22417 financed by Comunidad de Madrid and Universidad Complutense de Madrid. JR acknowledge financial support from the grants AYA2015-65973-C3-1-R and RTI2018-096228- B-C31 (MINECO/FEDER, UE), as well as from the State Agency for Research of the Spanish MCIU through the “Center of Excellence Severo Ochoa” award to the Instituto de Astrofísica de Andalucía (SEV-2017-0709).

\section{Data Availability}

The KCWI data presented are available via the Keck Observatory Archive (KOA): \url{https://www2.keck.hawaii.edu/koa/public/koa.php} 18 months after observations are taken. The JRT data can be acquired by contacting JSG through the email provided for correspondence. 




\bibliographystyle{mnras}
\bibliography{bibliography.bib} 




\appendix

\section{UDG Properties from the Literature}\label{app:summary}
In this appendix we provide a literature summary of UDGs with either stellar or GC velocity dispersions and the resulting dynamical masses for them using Equation \ref{eqtn:wolf}. For objects with both a GC and stellar velocity dispersion (i.e., NGC 5846\_UDG1, VCC 1287 and NGC1052-DF2) we quote only the stellar velocity dispersion. For objects with velocity dispersions from both MUSE ($\sigma_{inst}$ = 45 $\mathrm{km\ s^{-1}}$) and KCWI ($\sigma_{inst}$ = 13 $\mathrm{km\ s^{-1}}$) we give preference to the latter as we expect them to be more accurate due to the better instrumental resolution. In both cases literature works that derive the same parameters (e.g., velocity dispersion) for a UDG but are unused in the table are listed in the notes. 

Equation \ref{eqtn:wolf} was derived with the intent of using the luminosity weighted line-of-sight velocity dispersion as measured within 2D half-light radius \citep{Wolf2010}. Generally, this radius differs from that in which the velocity dispersion is actually measured. In order to calculate a dynamical mass from the \citet{vanDokkum2019} GC velocity dispersion we assume that their velocity dispersion measurement is equivalent to the line-of-sight velocity dispersion within half-light radius (i.e., a flat velocity profile for the galaxy). \citet{Toloba2018} used Equation \ref{eqtn:wolf} along with the GC velocity dispersion to calculate their dynamical masses. We note that, for some GC velocity dispersions the small number statistics can lead to both accuracy and precision loss in the measurement.

There is also some ambiguity as to whether or not all of the galaxies listed in Table \ref{tab:data_summary} should be treated equally when investigating the formation of UDGs. As such we caution the following:

\begin{itemize}
    \item Only 2 of the 9 UDGs in the \citet{Chilingarian2019} sample (those beginning with J in Table \ref{tab:data_summary}) meet the standard UDG definition with the other 7 being either slightly too small or slightly too bright.
    \item While Andromeda XIX meets the formal UDG definition it has orders of magnitude less stellar mass than other UDGs and is significantly fainter than what is realistically observable for non-Local Group UDGs.
    \item The UDGs NGC1052-DF2 and NGC1052-DF4 have both been inferred to have little to no dark matter and there exists debate in the literature as to their basic properties (e.g., their distance; \citealp{Trujillo2019, Monelli2019} although see \citealp{DanielieTRGB}). While both clearly fit the standard UDG definition, it is unclear how any single current proposed formation scenario could create both these UDGs and the other known UDGs (over-massive or otherwise). We therefore advise caution to those treating these galaxies as the other UDGs, seeking to find a singular formation mechanism to explain all. 
\end{itemize}

It should also be noted that \citet{Makarov2015} reports stellar velocity dispersions for three UDG-like galaxies. We choose not to include them in this compilation as they are measured at $\sim 10\%$ of the instrumental resolution and are listed as being highly uncertain.  

\begin{landscape}

\begin{table}
\caption{A summary of the UDG data plotted in Figures \ref{fig:sigma_mstar} and \ref{fig:ml}. \textbf{Bold} values indicate they have been calculated in this work. We use equation \ref{eqtn:wolf} to calculate dynamical masses for literature data. `--' denotes values missing. Errors are given in (brackets).}
\label{tab:data_summary}

\begin{tabular}{lllllllll}
\hline
Object &$R_{\mathrm{e, circ}}$ & $\sigma$ & $M(<r_{1/2})$ & $M_{\mathrm{*}}$ & M/L & Filter & References & Notes\\
 & [kpc] & [$\mathrm{km\ s^{-1}}$] & [$\times 10^8 M_\odot$] & [$\times 10^7 M_\odot$] & [$M_{\odot} / L_{\odot}$] &  & \\ \hline
 
Stellar velocity dispersions \\ \hline
UDG1137+16$^{\dag}$ &\textbf{3.9}(\textbf{0.1}) & \textbf{15} (\textbf{4}) & \textbf{6.18}  (\textbf{3.48}) & \textbf{14} (\textbf{2}) & \textbf{17.7} (\textbf{10.2}) & r & This work & 1 \\
VCC 1287$^{*}$ & 3.3 (--) & 19 (6) & 11.1 (8.1) & 20 (--) & 13 (11) & i & \citet{Pandya2018, Gannon2020} & 2\\
DGSAT I$^{\ddag}$ & \textbf{4.4} (\textbf{0.2}) & 56 (10) & 130 (50) & 27 (2) & 71 (25) & V & \citet{Martinez-Delgado2016, Martin-Navarro2019} & 3\\
Dragonfly 44$^{*}$ &\textbf{3.9} (--) & 33 (3) & 39 (5) & 30 (--) & 26 ($_{-6}^{+7}$) & $I_{814}$ & \citet{vanDokkum2016, vanDokkum2017, vanDokkum2019b} & 4 \\
DFX1$^{*}$ & \textbf{2.8} (--) & 30 (7) & \textbf{23} (\textbf{5}) & \textbf{34} (--) & \textbf{13} (\textbf{3}) & V & \citet{vanDokkum2017} & 1; 5; 6 \\
NGC 5846\_UDG1$^{\dag}$ & 2.14 (0.06) & 17 (2) & 5.75 (1.35) & 11 (--) & 20.9 (4.9) & V & \citet{Forbes2021} & 7; 8 \\
NGC1052-DF2$^{\dag}$ & 2 (--) & 8.5 ($_{-3.1}^{+2.3}$) & 1.3 (0.8) & 20 (--) & \textbf{1.3} (\textbf{1}) & V & \citet{Cohen2018, vanDokkum2018,Danieli2019} & 8; 9\\
Andromeda XIX$^{\dag}$ & 3 (1) & 7.8 ($_{-1.5}^{+1.7}$) & 1.1 (0.5) & 0.079 (--) & 278 ($_{-198}^{+146}$) & V & \citet{Martin2016, Collins2020} & 10\\
$\mathrm{J} 125846.94+281037.1^{*}$ & 1.3 (--) & 18 (13) & \textbf{3.92} (\textbf{5.66}) & \textbf{5.21} (--) & 16 (4) & R & \citet{Chilingarian2019} & 11; 12\\
$\mathrm{J} 125904.06+281422.4^{*}$ & 1.3 (--) & 25 (7) & \textbf{7.56} (\textbf{4.23}) & \textbf{7.02} (--) & 22 (8) & R & \citet{Chilingarian2019} & 11; 12\\
$\mathrm{J} 125904.20+281507.7^{*}$ & 1.2 (--) & 17 (10) & \textbf{3.23} (\textbf{3.79}) & \textbf{5.93} (--) & 12 (5) & R & \citet{Chilingarian2019} & 11; 12\\
$\mathrm{J} 125929.89+274303.0^{*}$ & 2.1 (--) & 21 (7) & \textbf{8.61} (\textbf{5.74}) & \textbf{11.19} (--) & 4.5 (2.5) & R & \citet{Chilingarian2019} & 11\\
$\mathrm{J} 125937.23+274815.2^{*}$ & 0.9 (--) & 22 (8) & \textbf{4.05} (\textbf{2.95}) & \textbf{6.81} (--) & 22 (10) & R & \citet{Chilingarian2019} & 11; 12 \\
$\mathrm{J} 130005.40+275333.0^{*}$ & 1.4 (--) & 37 (6) & \textbf{17.82} (\textbf{5.78}) & \textbf{6.84} (--) & 47 (17) & R & \citet{Chilingarian2019} & 11; 12\\
$\mathrm{J} 130026.26+272735.2^{*}$ & 3.7 (--) & 19 (5) & \textbf{12.42} (\textbf{6.54}) & \textbf{15.71} (--) & 2 (0.6) & R & \citet{Chilingarian2019} & 11 \\
$\mathrm{J} 130028.34+274820.5^{*}$ & 1.3 (--) & 23 (7) & \textbf{6.4} (\textbf{3.89}) & \textbf{4.85} (--) & 22 (6) & R & \citet{Chilingarian2019} & 11; 11 \\
$\mathrm{J} 130038.63+272835.3^{*}$ & 1.9 (--) & 27 (5) & \textbf{12.88} (\textbf{4.77}) & \textbf{7.32} (--) & 11 (4) & R & \citet{Chilingarian2019} & 11; 13\\ \hline

GC velocity dispersions\\ \hline
VLSB-B$^{*}$ & 2.9 (0.2) & 47 ($_{-29} ^{+53}$) & 49 ($_{-49}^{+111}$) & 0.6 (0.1) & 407 ($_{-407}^{+916}$) & V & \citet{Toloba2018} & 14; 15 \\
VLSB-D$^{*}$ & 13.4 (2) & 16 ($_{-4}^{+6}$) & 32 ($_{-17}^{+24}$) & 7.9 (0.1) & 21 ($_{-11}^{+15}$) & V & \citet{Toloba2018} & 14; 15\\
VCC 615$^{*}$ & 2.4 (0.1) & 32 ($_{-10}^{+17}$) & 25 ($_{-1.6}^{+2.7}$) & 2.1 (0.1) & 60 ($_{-38}^{+65}$) & V & \citet{Toloba2018} & 14; 15\\
NGC1052-DF4$^{\dag}$ & \textbf{1.5} (--) & 4.2 ($_{-2.2}^{+4.4}$) & \textbf{0.25} (\textbf{$_{-0.26}^{+1.04}$}) & 15 (4) & 0.64 ($_{-0.74}^{+2.96}$) & $V_{606}$ & \citet{vanDokkum2019} & 8; 15; 16\\
\hline
\end{tabular}
\end{table}
Notes.---

\begin{multicols}{2}
\begin{enumerate}[label=\arabic*)]
    \item A $M_{\mathrm{*}}/L = 2$ was used to calculate the stellar mass. 
    \item See also \citet{Beasley2016}.
    \item $R_{e}$ circularised using literature b/a (0.87 - \citealp{Martinez-Delgado2016}).
    \item $R_{e}$ circularised using literature b/a (0.68 - \citealp{vanDokkum2017}).
    \item $R_{e}$ circularised using literature b/a (0.62 - \citealp{vanDokkum2017}).
    \item It is unclear if the velocity dispersion reported in \citet{vanDokkum2017} for DFX1 is also effected by the same problem reported in \citet{vanDokkum2019b} for Dragonfly 44 from the same work.
    \item See also \citet{Muller2020} who dubNGC 5846\_UDG1, MATLAS-2019. 
    \item $M_{\mathrm{*}}/L = 2$ was assumed in the literature to calculate stellar mass.
    \item See also work by \citet{vanDokkum2018}, \citet{Ruiz-lara2019} and \citet{Emsellem2018} for NGC1052-DF2.
    \item Dynamical mass was calculated using the slightly different \citet{Walker2009} mass estimator instead of that from \citet{Wolf2010}. 
    \item Stellar mass calculated using the \citet{Chilingarian2019} stellar M/L ratio listed in their table 1.
    \item Too small for standard UDG definition ($R_{e} > 1.5$ kpc).
    \item Too bright for standard UDG definition ($\mu_{0,g} > 24\ \mathrm{mag\ arcsec^{-2}}$).
    \item Dynamical mass is calculated using Equation \ref{eqtn:wolf} with $R_{e}$ as the radius containing half the number of GCs.
    \item Dynamical mass calculation assumes GC velocity dispersion = stellar velocity dispersion.
    \item $R_{e}$ circularised using literature b/a (0.89 - \citealp{vanDokkum2019}).\\
    $^{*}$ Cluster UDG \quad $^{\dag}$ Group UDG \quad $^{\ddag}$ Field UDG
\end{enumerate}
\end{multicols}


\bsp	
\label{lastpage}

\end{landscape}

\end{document}